\documentclass[10pt, twocolumn]{IEEEtran}
\usepackage{spconf, epsfig, amsmath, amssymb}

\begin{document}

\setlength{\evensidemargin}{-0.35in}
\setlength{\oddsidemargin}{-0.35in}

\setlength{\textheight}{9in}

\setlength{\topmargin}{0in}
\setlength{\headheight}{0in}
\setlength{\headsep}{0in}


\title{\Large\bfseries Maximizing compression efficiency through block rotation}
%
\name{Rui F. C. Guerreiro\thanks{\noindent Partially supported by FCT, under ISR/IST plurianual funding (POSC program, FEDER) and grants MODI-PTDC/EEA-ACR/72201/2006 and SFRH/BD/48602/2008.} \hspace*{2cm} Pedro M. Q. Aguiar}
\address{Institute for Systems and Robotics, Instituto Superior T\'{e}cnico\\
 Av. Rovisco Pais, 1049-001 Lisboa, Portugal\\
{\tt ruifcguerreiro@hotmail.com}, \hspace*{.1cm} {\tt aguiar@isr.ist.utl.pt} }


%
%
%
%
\maketitle
\begin{abstract}
The Discrete Cosine Transform (DCT) is widely used in lossy image and video compression schemes, \it e.g.\rm, JPEG and MPEG. In this paper, we show that the compression efficiency of the DCT is dependent on the edge directions within a block. In particular, higher compression ratios are achieved when edges are aligned with the image axes. To maximize compression for general images, we propose a \it rotated block DCT method\rm. It consists of rotating each block, before applying the DCT, by an angle that aligns the edges, and rotating back the block in the decompression stage. We show how to compute the rotation angle and analyze two alternative block rotation approaches. Our experiments show that our method enables both a perceptual improvement and a PSNR increase of up to 2dB, compared with the standard DCT, for low and medium bit rates.
\end{abstract}


%
\begin{keywords}
Discrete Cosine Transforms, image coding
\end{keywords}
\section{Introduction}
\label{sec:intro}

\noindent JPEG~\cite{JPEGbook} is a commonly used method for lossy compression of natural images. It encodes images using fewer bits than an unencoded representation would, by discarding data that is less perceptible to the eye. Studies have shown that natural images have a low amount of sudden intensity changes~\cite{naturalImages_94} and perceptual studies show that the Human Visual System is not very sensitive to this data either~\cite{HumanVisualSystem_Bouman67}. Therefore, the main idea behind JPEG is the removal of high-frequency data by using the two-dimensional DCT as an efficient way to break up the underlying structure of an image into different frequency coefficients. 

The main steps of JPEG are: 1) dividing the image in blocks of $8\times 8$ pixels; 2) applying the DCT to each block; 3) lossy quantization of DCT coefficients, according to a quantization matrix which accounts for their perceptual importance; 4) storing non-zero values. Most of the discarded bits and coefficients are the high-frequency ones. The decompression stage creates a decompressed image from the stored coefficients and the Inverse DCT (IDCT).

In this paper, we show that the DCT encodes horizontal and vertical edges with fewer coefficients than slanted ones. This shows that the compression efficiency enabled by the DCT is dependent on the edge directions within each block, and that quality improvements are possible if this fact is exploited. We propose a \it rotated block DCT method \rm that rotates each block, before applying the DCT, by an angle that maximizes efficiency, and rotates the block back in the decompression stage. We show how to compute the rotation angle and introduce two alternative rotation approaches. Because a rotated block is a rhombus and the square block that includes it has missing data, we also discuss how to estimate it. We test our method in a simplified compression scheme and show that it enables perceptual improvements and a \it Peak Signal-to-Noise Ratio \rm (PSNR) increase of up to 2dB compared with the standard DCT, for low and medium bit rates.

\noindent \bf Paper organization \rm Section~\ref{sec:section2} studies the dependence of the DCT compression efficiency on the block content. Section~\ref{sec:section3} proposes the rotated block DCT method. Section~\ref{sec:experiments} shows experimental results and we conclude on section~\ref{sec:conclusions}.

\section{Compression efficiency of the 2D DCT}
\label{sec:section2}

\begin{figure}[htb]
\begin{minipage}[b]{1.0\linewidth}
\centering
 \centerline{\includegraphics[width=0.7\linewidth]{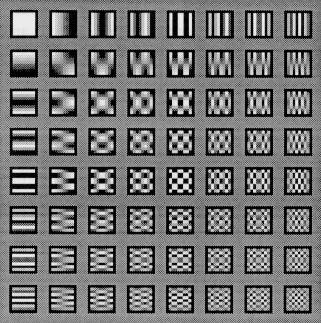}}
\end{minipage}

\vspace*{-.3cm}
\caption{DCT basis functions, for $N = 8$}\label{fig:DCT}
\vspace*{-0.35cm}
\end{figure}

\noindent The DCT~\cite{DIPGonzalez} is a Fourier-related frequency transform with real frequency coefficients, given by
\begin{align*}
&\mathcal{F}\left[I\right]_{ij} = \displaystyle\sum_{x = 0}^{N-1} \sum_{y = 0}^{N-1} I\left(x,y\right)B_{ij}\left(x,y\right),
\\
B_{ij}\left(x, y\right) &= \cos\left[\frac{\pi}{N}\left(x + \frac{1}{2}\right)i\right]\cos\left[\frac{\pi}{N}\left(y + \frac{1}{2}\right)j\right],
\end{align*}
where $I$ is an $N\times N$ block, $\mathcal{F}\left[I\right]$ are the corresponding frequency coefficients and $\left\{B_{ij}\right\}$ are Fourier-related basis functions, graphically illustrated in Figure~\ref{fig:DCT}. 

Since the DCT basis include pure horizontal and vertical cosine functions (top row and left column of Figure~\ref{fig:DCT}), edges aligned with the image axes are encoded with a small number of DCT coefficients. This enables high compression efficiency. Slanted edges, on the other hand, require a larger amount of DCT coefficients, due to the inadaptation to the basis functions of the DCT, in Figure~\ref{fig:DCT}. This reduces compression efficiency.



\begin{figure}[htb]
\begin{minipage}[b]{0.32\linewidth}
  \centering
 \centerline{\includegraphics[width=1.0\linewidth]{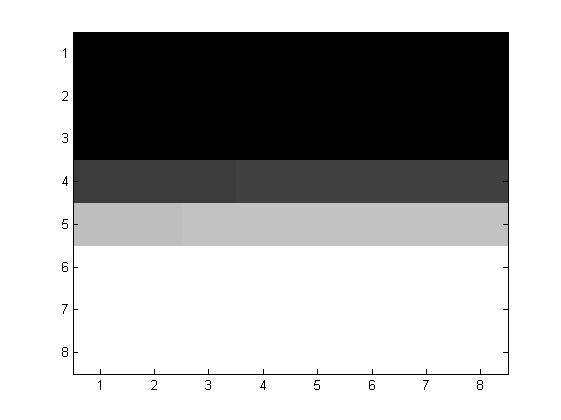}}
 \centerline{\includegraphics[width=1.0\linewidth]{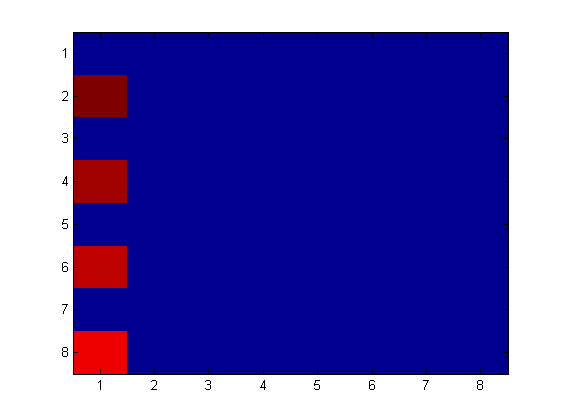}}
  \vspace{-0.2cm}
  \centerline{(a) $\theta = 0^{\circ}$}\medskip
\end{minipage}
\hfill
\begin{minipage}[b]{0.32\linewidth}
  \centering
 \centerline{\includegraphics[width=1.0\linewidth]{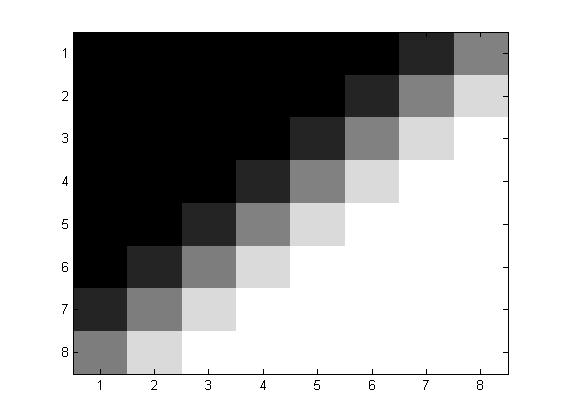}}
 \centerline{\includegraphics[width=1.0\linewidth]{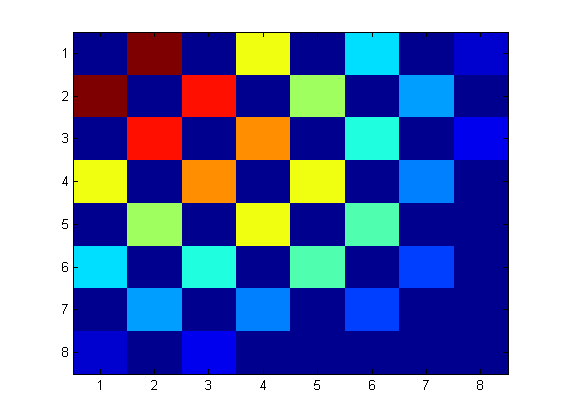}}
  \vspace{-0.2cm}
  \centerline{(b) $\theta = 45^{\circ}$}\medskip
\end{minipage}
\hfill
\begin{minipage}[b]{0.32\linewidth}
  \centering
 \centerline{\includegraphics[width=1.0\linewidth]{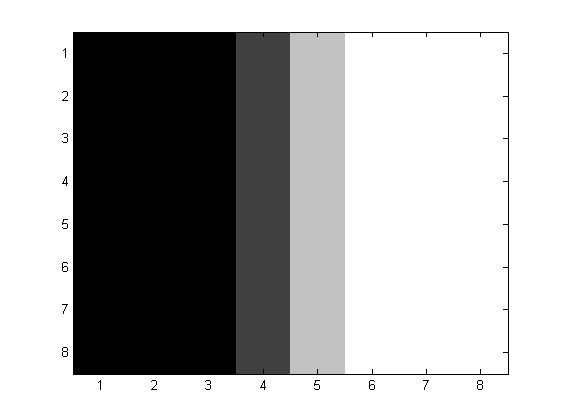}}
 \centerline{\includegraphics[width=1.0\linewidth]{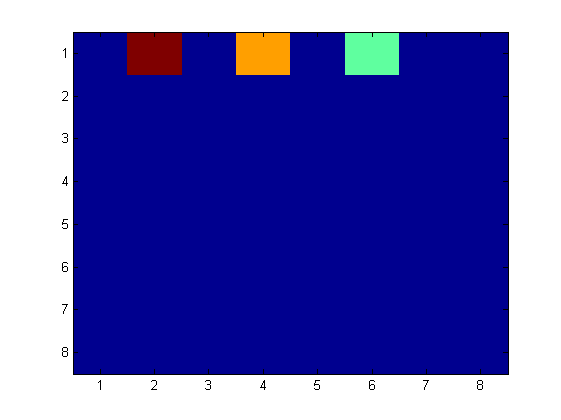}}
  \vspace{-0.2cm}
  \centerline{(c) $\theta = 90^{\circ}$}\medskip
\end{minipage}
\vspace*{-.5cm}
\caption{Top row: step edge, $I_{\theta}$. Bottom row:  $\log\left(\left|\mathcal{F}\left[I_{\theta}\right]\right|\right)$}\label{fig:DCTefficiency}
\vspace*{-.29cm}
\end{figure}

We now illustrate this effect using a step edge block, $I_{\theta}$, shown on the top row of Figure~\ref{fig:DCTefficiency}, for $\theta = \left\{0^{\circ}, 45^{\circ}, 90^{\circ}\right\}$. The bottom row of Figure~\ref{fig:DCTefficiency} shows the logarithm of the magnitude of the DCT coefficients of $I_{\theta}$, $\log\left(\left|\mathcal{F}\left[I_{\theta}\right]\right|\right)$, confirming that horizontal and vertical edges are coded with fewer coefficients than slanted ones. In general, this illustrates that the edge directions within a block affect the compression efficiency of the DCT, and that quality improvements are possible by exploring this fact.


\section{Rotated block DCT method}
\label{sec:section3}

\noindent In this section, we propose the rotated block DCT method, as an alternative to the standard DCT. This methods consists of, for each block, estimating the rotation angle, $\hat{\theta}$, that maximizes the DCT compression efficiency and creating an \it extended block\rm, $I_{\theta}$, containing the rotated block. It estimates the missing values of the extended block and applies the DCT, to obtain frequency coefficients. To decompress the image, coefficients $\tilde{I}_{\theta}$ are decoded using the IDCT and rotated back to their original angle. 



%
%
%

\subsection{Angle estimation}
\label{ssec:angleEstimation}

\noindent The angle by which we rotate a block, $\hat{\theta}_n$, is the one that minimizes the \it Mean Square Error \rm between the original block and its reconstruction, $R_{\theta n}$. The reconstruction is obtained by rotating a block by the angle $\theta$, computing its DCT, taking the $n$ largest coefficients, computing the IDCT of these numbers, and rotating the block back. Formally,
\begin{align}
&\hat{\theta}_n =\arg \min_{\theta}  \displaystyle\frac{1}{64}\sum_{x = 0}^{7}\sum_{y=0}^{7} \left\| I(x,y)- R_{\theta n}(x,y)\right\|_2^2,\label{eq:angleEstimation}
\\
&R_{\theta n} = \mathcal{F}^{-1}\left[L_n\left(\mathcal{F}\left[I_{\theta}\right]\right)\right]_{-\theta},\label{eq:reconstruction}
\end{align}
where $L_n$ keeps the $n$ largest coefficients and zeroes the remaining ones. $\hat{\theta}$ is obtained by exhaustive search, for $\theta\! \in \!\left[0^{\circ},90^{\circ}\right[$. The computation of $I_{\theta}$ is detailed in the sequel.


\subsection{Block rotation}
\label{ssec:blockRotation}

\noindent A rotated square block is a rhombus, for which the DCT is not defined. To apply the DCT, we define an \it extended block\rm, $I_{\theta}$, a square block which contains the entire rhombus. We rotate each block using Bicubic interpolation~\cite{bicubicInterpolation81}. We introduce two alternative rotation approaches.

\begin{figure}[htb]
\begin{minipage}[b]{1.0\linewidth}
\centering
 \centerline{\includegraphics[width=0.5\linewidth]{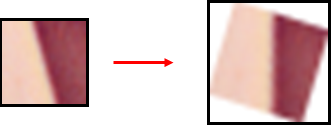}}
\end{minipage}
\vspace*{-.7cm}
\caption{Extended block, with constant sampling rate}\label{fig:upscaleIllustration}
\vspace*{-.15cm}
\end{figure}

\noindent \bf Constant sampling rate \rm Preserving the sampling rate, rotating an $8 \times 8$ block originates an extended block that can reach up to $12 \times 12$, as illustrated in Figure~\ref{fig:upscaleIllustration}. This has the advantage of preserving high-frequency data but requires DCTs of various sizes, which increases complexity and the amount of frequency coefficients to be compressed.



\begin{figure}[htb]
\begin{minipage}[b]{1.0\linewidth}
\centering
 \centerline{\includegraphics[width=0.5\linewidth]{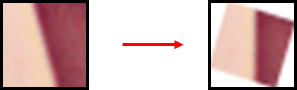}}
\end{minipage}
\vspace*{-.7cm}
\caption{Extended block, with constant block size}\label{fig:downscaleIllustration}
\vspace*{-.6cm}
\end{figure}

\noindent \bf Constant block size \rm Constraining the extended block to have constant $8\times 8$ size has the advantage of using a single DCT and a low amount of coefficients. However, as illustrated in Figure~\ref{fig:downscaleIllustration}, the sampling rate may diminish up to $30\%$, which removes high-frequency data. Since the sampling rate is angle-dependent, the compression efficiency advantage of rotating a block is somewhat counterbalanced by the high-frequency data removal. A constant extended block of size $12\times 12$ preserves high-frequency data but produces a large amount of frequency coefficients.

\subsection{Missing data estimation}
\label{ssec:missingData}


\noindent As illustrated in Figures~\ref{fig:upscaleIllustration} and~\ref{fig:downscaleIllustration}, the extended block contains missing entries. These entries may be set to a constant value, or they can be obtained from the neighboring values of the original $8\times 8$ block. Setting these entries to a constant value may originate artificial edges within the extended block, which are harder to encode and thus reduce compression efficiency. Using the neighboring values of the original block, \it i.e\rm, rotating a rhombus of the original image to fit the complete extended block, is a natural way to solve this task. However, because the extra pixels might contain new edges and detail, the compression efficiency of the DCT is not optimal. We use the latter approach.

The missing entries could also be estimated by solving an optimization problem that maximizes the compression efficiency of DCT. However, this greatly increases computational cost and leads to minimal quality improvements.



\subsection{Approximate angle estimation}
\label{ssec:approximateAngleEstimation}

\noindent Although we compute $\hat{\theta}$ using~\eqref{eq:angleEstimation}, a computationally simpler way to compute $\hat{\theta}$ can also be obtained by estimating the predominant edge direction. Computing
\begin{align*}
\tilde{\theta}_{xy} = \arctan\left(\frac{\Delta_y I_{xy}}{\Delta_x I_{xy}}\right)+90k,
\end{align*}
where $\tilde{\theta}_{xy}$ is the angle estimated at position $(x,y)$, in degrees, $\Delta_xI_{xy}$ and $\Delta_yI_{xy}$ are the corresponding horizontal and vertical derivatives, and $k \in \mathbb{Z}$ forces $\tilde{\theta}_{xy}$ to lie in the interval $[0, 90^{\circ}[$. A histogram with 90 bins, $H$, would be created and computed using,
\begin{align*}
H_{\tilde{\theta}_{xy}} = H_{\tilde{\theta}_{xy}} + \left\|\Delta I_{xy}\right\|_2,
\end{align*}
for all $(x,y) \in \{0,...,7\}\times\{0,...,7\}$, where $\left\|\Delta I_{xy}\right\|_2 = \sqrt{\Delta_x I_{xy}^2+\Delta_y I_{xy}^2}$ . The estimate $\hat{\theta}$ would be the number of the largest bin of the (possibly smoothed) histogram.



\section{Experiments}
\label{sec:experiments}

\begin{figure}[htb]
\begin{minipage}[b]{1.0\linewidth}
\centering
 \centerline{\includegraphics[width=\linewidth]{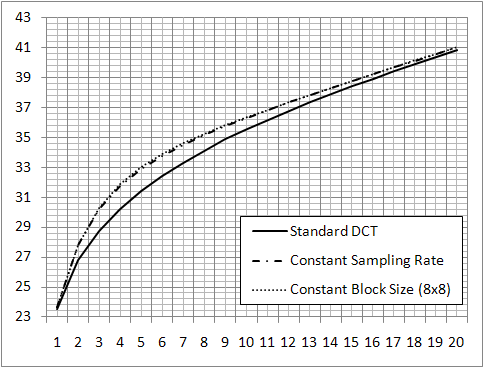}}
\end{minipage}
\vspace*{-.7cm}
\caption{PSNR/coefficient}\label{fig:PSNR_per_coefficient}
\end{figure}

\noindent We use the well-known Lenna image and a simple but illustrative quantization scheme, which preserves the $n$ largest DCT coefficients and zeroes the remaining ones. We chose to do so because the large signal chain of the JPEG compression scheme would make it harder to draw clear conclusions about our method. Figure~\ref{fig:PSNR_per_coefficient} shows the PSNR between the original image and the reconstructed images, as a function of the number of coefficients per block. The reconstructed images are obtained using~\eqref{eq:reconstruction} in each block, for the standard DCT (\it i.e.\rm, $\theta = 0$) and our method, with constant sampling rate and constant block size. Because only a few bits are required to represent the angle, it is not considered a coefficient in our analysis.


Figure~\ref{fig:PSNR_per_coefficient} shows that our method achieves a PSNR increase of up to 2dB compared with the standard DCT, for a low number of coefficients. This improvement confirms the main idea behind this paper, \it i.e.\rm, that adapting the content of each block to the basis functions of the DCT leads to more descriptive coefficients and higher quality reconstructions. Because the first coefficients serve, essentially, the purpose of reconstructing the main structure of each block, high-frequency data does not play a significant role. For this reason, methods with constant sampling rate and constant block size have similar performance.

Figure~\ref{fig:PSNR_per_coefficient} also shows that the advantage of using our method decreases, for larger numbers of coefficients. This occurs because high-frequency data plays a more significant role, which is important for two reasons. First, the rotation algorithms introduce important errors, mostly in high-frequency data, which now become apparent. This is especially true for a constant block size. Secondly, because high-frequency data, \it i.e.\rm, detail, has typically multiple directions, there is no longer a single rotation angle that can produce large compression efficiency improvements. For these reasons, the angle estimator~\eqref{eq:angleEstimation} increasingly chooses not to rotate the blocks and, therefore, performance approximates that of the standard DCT.

\begin{figure}[htb]
\begin{minipage}[b]{1.0\linewidth}
\centering
 \centerline{\includegraphics[width=\linewidth]{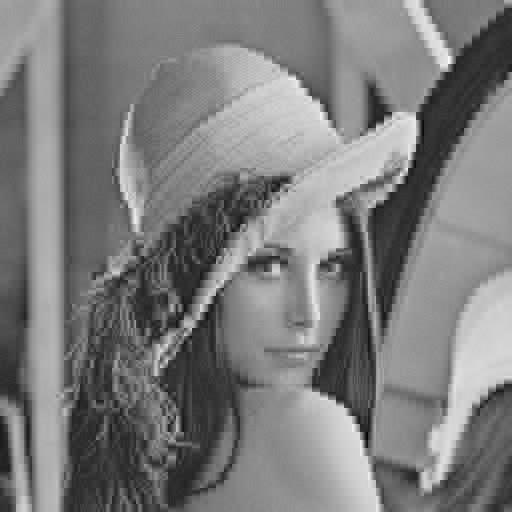}}
  \centerline{(a) standard DCT}\medskip
\end{minipage}
\vspace*{-.5cm}
\begin{minipage}[b]{1.0\linewidth}
  \centering
 \centerline{\includegraphics[width=\linewidth]{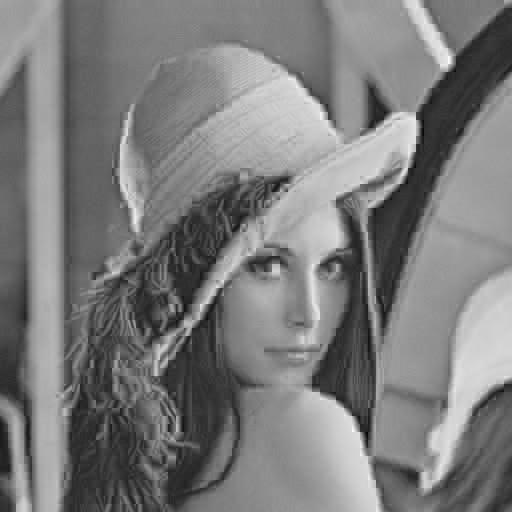}}
  \centerline{(b) our method, with constant sampling rate}\medskip
\end{minipage}
\vspace*{-.5cm}
\caption{Lenna, with 2 coefficients/block}\label{fig:Lenna10Coeffs}
\end{figure}
\begin{figure}[htb]
\begin{minipage}[b]{1.0\linewidth}
\centering
 \centerline{\includegraphics[width=\linewidth]{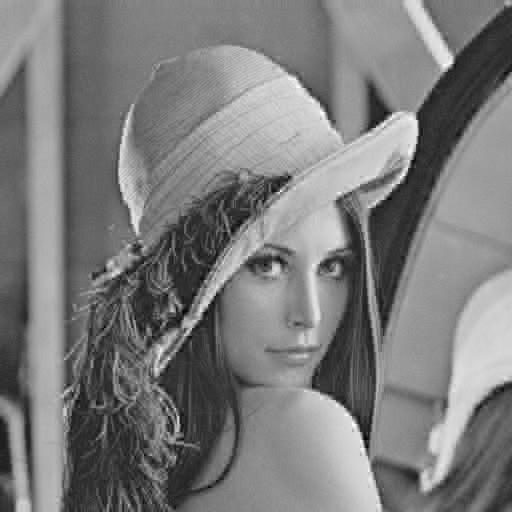}}
  \centerline{(a) standard DCT}\medskip
\end{minipage}
\vspace*{-.5cm}
\begin{minipage}[b]{1.0\linewidth}
  \centering
 \centerline{\includegraphics[width=\linewidth]{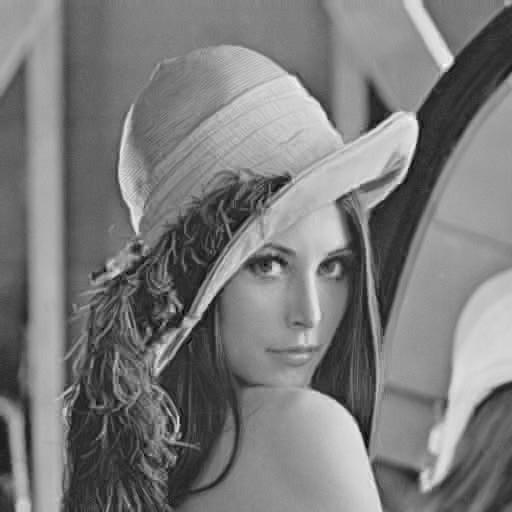}}
  \centerline{(b) our method, with constant sampling rate}\medskip
\end{minipage}
\vspace*{-.5cm}
\caption{Lenna, with 4 coefficients/block}\label{fig:Lenna15Coeffs}
\end{figure}

Because our method achieves significant quality improvements for PSNR values up to 38dB, it can play a significant role in low and medium bit-rate image compression applications. Figures~\ref{fig:Lenna10Coeffs} and~\ref{fig:Lenna15Coeffs} show the results obtained for the standard DCT and our method, for 2 and 4 coefficients, respectively. They show a clear perceptual advantage of using our method, compared with the standard DCT.


\section{Conclusions}
\label{sec:conclusions}

\noindent In this paper, we show that the compression efficiency of the DCT is dependent on the edge directions within a block. We propose the rotated block DCT method, which exploits this fact by rotating each block before applying the DCT, optimizing compression efficiency. We test our method on the Lenna image, using a simplified compression scheme. Our tests show a clear perceptual improvement and a PSNR increase of up to 2 dB, compared with the standard DCT, for low and medium bit-rates.

\bibliographystyle{IEEEbib}
\bibliography{references}

\end{document}